# Spectroscopy of metal "superatom" nanoclusters and high-$T_c$ superconducting pairing

Avik Halder and Vitaly V. Kresin

Department of Physics and Astronomy,

University of Southern California,

Los Angeles, CA 90089-0484, USA

*e-mail: kresin@usc.edu

A unique property of metal nanoclusters is the "superatom" shell structure of their delocalized electrons. The electronic shell levels are highly degenerate and therefore represent sharp peaks in the density of states. This can enable exceptionally strong electron pairing in certain clusters composed of tens to hundreds of atoms. In a finite system, such as a free nanocluster or a nucleus, pairing is observed most clearly via its effect on the energy spectrum of the constituent fermions. Accordingly, we performed a photoionization spectroscopy study of size-resolved aluminum nanoclusters and observed a rapid rise of the near-threshold density of states of several clusters ($Al_{37,44,66,68}$) with decreasing temperature. The characteristics of this behavior are consistent with compression of the density of states by a pairing transition into a high-temperature superconducting state with $T_c \gtrsim 100$ K. This value exceeds that of bulk aluminum by two orders of magnitude. These results highlight the potential of novel pairing effects in size-quantized systems and the possibility to attain even higher critical temperatures by optimizing the particles' size and composition. As a new class of high-temperature superconductors, such metal nanocluster particles are promising building blocks for high-$T_c$ materials, devices, and networks.

# 1 Introduction

Size effects in superconductivity – changes occurring when one or more dimensions of the sample become so small as to exhibit significant quantum effects due to the confinement of electrons – have long been of interest to researchers. There is obvious value in pursuing novel systems that can support higher critical temperatures, critical currents and critical magnetic fields, or display other beneficial and unusual properties. Nanoscale-based materials in general, and in particular those in which the size and composition of the constituent building blocks can be accurately manipulated, represent an especially interesting and fruitful realm for this exploration.

The study of size-selected metal clusters, also known as nanoclusters [1-3], focuses on precisely this target: by mapping out the evolution of metal properties with size, one can observe and select the system of interest with atomic precision. One of the most remarkable quantum size effects in nanoscience is the electronic shell structure displayed by such clusters (see, for instance, the review [4]). For many materials which become good conductors in the bulk, the future conduction electrons become delocalized even in a small particle and occupy discrete energy levels which organize into clear shell ordering, akin to that in the periodic table or in nuclei. As in these cases, the electronic states in nanoclusters can be characterized by their angular momentum quantum number $l$. The high stability associated with shell filling was originally discovered via increased abundance of corresponding cluster sizes monitored in a molecular-beam experiment [5]. The existence of such shell structure has been directly proven by photoelectron spectroscopy (see, e.g., the reviews [6,7]). Clusters displaying shell ordering of their delocalized electrons' energy levels are often referred to as "superatoms" [8].

In the context of superconductivity, the presence of quantum shell structure can lead to dramatic implications for electron pairing. Closed-shell spherical ("magic") clusters have a level degeneracy of $2(2l+1)$. For example, one of the clusters discussed below, $Al_{66}$, fits 30 electrons into its relatively narrow $1j$ highest occupied shell [9,10]. Qualitatively, the shell degeneracy in a cluster can be viewed as a sharp peak in the density of states (DOS) near the Fermi level, akin to a Van Hove singularity [11]. This amplifies the pairing coupling constant $\lambda$, which is proportional to the DOS, and greatly enhances the gap parameter $\Delta$ and the critical temperature. In some (but by no means all) cluster sizes, a propitious combination of a large $\lambda$ and an appropriate intershell spacing $\delta\varepsilon$ can create a situation favorable for very high $T_c$ [13,14].

Formation of a superconducting state in finite Fermi systems is in fact a recognized phenomenon. One well-known analogy to electrons in clusters is the atomic nucleus, where pairing was surmised almost immediately after the appearance of BCS theory [15-17]. In both cases the pairs are composed of fermions with opposite projections of orbital and spin angular momenta. What's more, in both cases the formation of Coper pairs is distinctly manifested by its effect on the energy spectrum of the system. Pairing is also actively explored in trapped atomic gas clouds (see, e.g., [18,19]), and has been discussed for conjugated organic molecules (e.g., [14,20,21]).

Strengthening of superconductivity in finite metal grains and nanoparticles, driven by size quantization, has been studied for many years (for example [22-25]; see the reviews in [14,26-29]), with reported $T_c$ enhancement by factors as large as ~2-3. The enhancement predicted for nanoclusters with shell structure, on the other hand, can reach as much as 2 orders of magnitude (thanks to the aforementioned high orbital degeneracy).

The fact that nanocluster shell structure is promising for superconductivity was already noted by such authors as J. Friedel [30], W. D. Knight [31], and B. Mottelson [32] but the detailed theoretical analysis and its quantitative prediction of great strengthening relative to the bulk appeared more recently, as cited above. This rigorous treatment employs the strong-coupling formalism and incorporates into it both the discrete nature of the electrons' spectrum and the conservation of their number, as appropriate for a finite Fermi system. It is also verified that fluctuations of the order parameter will broaden but not destroy the pairing transition: thanks to the large values of $T_c$ and $\Delta$, the coherence length becomes small, i.e., comparable to the cluster size, hence in this respect the system is not zero-dimensional.

The publication [13] was followed by a number of calculations by different groups [14,33-35]. Thus the prediction of high-temperature superconducting pairing in individual size-selected clusters with shell structure rests on solid theoretical foundation and is ripe for experimental verification.

What is nontrivial, of course, is how to probe for the appearance of pairing correlations in individual clusters flying in a molecular beam. Temperature control is not as straightforward as in a cryostat, but this can be handled by proper source design (see below). More fundamentally, one cannot do a resistance measurement, and the Meissner effect would be too weak for magnetometry or Stern-Gerlach-type [36] beam deflection [37]. (In addition, a Larmor diamagnetic response would be exhibited by closed-shell clusters even in their normal state [40-42].)

The solution to this experimental challenge is the aforementioned fact that pairing has spectroscopically observable consequences. The appearance of a gap modifies the excitation spectrum, and this can be detected by a careful measurement. In this respect, the situation is parallel to the detection of superconducting correlations in atomic nuclei [15-17].

The technique applied in the present work is size-selective photoionization spectroscopy on a thermalized cluster beam: an optimized source generates a beam of clusters at a defined temperature, a tunable laser ionizes the clusters and produces a map of their electrons' DOS, and a mass spectrometer sorts the clusters by size. An earlier brief report was published in Ref. [43], here we provide a full description of the experimental procedure, and present further results and a detailed discussion.

The plan of the paper is as follows. In Sec. 2 we describe the experimental apparatus and procedure. Sec. 3 contains a detailed discussion of the data on the closed-shell "magic" superatom cluster $Al_{66}$. Sec. 4 describes the data on three "non-magic" nanoclusters, and Sec. 5 offers a summary and comments about further work.

## 2 Experiment

Optimal candidate materials for the exploration of nanocluster pairing should satisfy two conditions. It is favorable if they are superconductors in the bulk state, so as to provide confidence that the electron-vibrational coupling is sufficiently strong. They should also be known to display shell structure in nanocluster form. Among the possibilities are Al, Zn, Cd, Ga and In. We chose to work with aluminum because it is a well-known superconductor (crystalline $T_c$=1.2 K, amorphous $T_c$=6 K [44]) and at the same time many $Al_n$ clusters with $n>40$ are well described by the shell model [9,10,45-48]. In addition, the metal is essentially isotopically pure $^{27}$Al which eliminates any complications with mass spectrometric identification of cluster sizes.

Fig. 1 shows an outline of the experiment. Neutral clusters are formed inside a homebuilt magnetron sputtering/condensation source based on the design described in [49,50]. Metal vapor is produced from a 1 inch diameter target by argon ion sputtering (Ar inlet flow rate 100 sccm, discharge voltage 250 V, discharge power 40 W). A continuous flow of helium gas is also fed into the chamber, at a rate approximately three times that of argon. The gas mixture entraps the sputtered metal atoms and carries them, at a pressure of ≈0.8 mbar, through the 10 cm long aggregation region (a 7.6 cm diameter liquid nitrogen cooled tube) where cluster nucleation takes place.

As mentioned above, the ability to adjust the temperature of the clusters in the beam is an essential part of the experiment. To enable this, we equipped the magnetron source with a "thermalizing tube" which attaches directly to the exit hole of the aggregation region. By extensive trials, the following dimensions were found to offer a satisfactory combination of particle flux, size distribution, and collimation at all beam temperatures: length of 12 cm, inner diameters of 16 mm and 8 mm for $T$ above and below 90 K, respectively, and a 6 mm diameter exit aperture. The gas flow conditions listed above result in a pressure of ≈0.6 mbar inside this tube, which ensures that the clusters undergo at least ~$10^5$ collisions with the buffer gas and equilibrate with the tube wall temperature to within ±1 K [51,52]. The tube was machined out of oxygen-free high conductivity copper with 12.5 mm walls for increased thermal conductivity. Over the studied temperature range of 65 K – 230 K the inner surface was equilibrated within ±1 K along its full length, as monitored by platinum resistance temperature detectors (above 90 K) or silicon-diode sensors (below 90 K; all sensors from Omega Engineering) embedded deep inside the wall. For temperatures down to 90 K, the temperature was adjusted by balancing good thermal contact with the liquid nitrogen-cooled aggregation chamber using counterheating with a band electric heater. For lower temperatures, the thermalizing tube was isolated from the aggregation chamber by a teflon spacer and connected with the first stage of a closed-cycle helium refrigerator (CTI Cryogenics Model 22) by strands of thick silver-coated copper braid. In all cases, the tube was surrounded by multiple layers of superinsulation.

Nanoclusters exiting the thermalizing tube pass a 2 mm-diameter conical skimmer positioned 2 cm away. The source chamber pressure is maintained at ≈3·$10^{-3}$ mbar by a Varian VHS-10 pump with an extended cold cap.

Downstream, the clusters are ionized by 5 ns pulses from a tunable Nd:YAG/OPO laser system (EKSPLA NT342/3/UV). The laser fluence $\Phi$ is attenuated by a neutral density filter and maintained at ~500 μJ/cm$^2$ to ensure single photon absorption, as verified by the linearity of the ion yield $Y(\Phi)$ [53]. The ionization takes place within the homebuilt extraction region of a linear Wiley-McLaren time-of-flight (TOF) mass spectrometer followed by a 1.3 m flight path to a channeltron ion detector. (Fine mesh coverings on TOF plate apertures were very helpful for reducing divergence of the extracted ion beam.) The custom-built channeltron (DeTech Inc.) contains a conversion dynode which can be operated at up to 20 kV (the present measurement used 14 kV), which dramatically enhances the efficiency of detecting heavy cluster ions. Time-of-flight mass spectra are collected using a multichannel scaler (ORTEC MCS-pci).

In Fig. 2 we show one of the time-of-flight mass spectra; their shape at a given wavelength remains qualitatively the same at all temperatures. In deconvoluting the mass spectrum, we found that each peak may overlap at most with its second nearest-neighbor. Hence for each cluster size Al$_x$ the intensity was found by fitting five Gaussians to the points ranging from Al$_{x-2}$ to Al$_{x+2}$ and then integrating the strength of the central peak. An example is shown in the inset in Fig. 2.

The ion yield values, $Y(\hbar\omega)$, must be normalized to account for the intensity variations of the light pulses and for the possible drift of the cluster beam flux. The former is accomplished by constantly recording the laser pulse energy immediately past the ionization region, while the latter is taken into account by normalizing all measured ion rates to reference spectra taken at 216 nm after each collection interval. The data were acquired in the wavelength range 210-250 nm in steps of 1 nm at five different temperatures: 65K, 90K, 120K, 170K and 230K. Each measurement for a given temperature lasted ~25-30 hours and was repeated 3-5 times. This long collection time helped to nullify intensity fluctuations as well as to enhance system stability and data statistics. At the same time, it limited the number of temperature points which could realistically be mapped out in the experiment.

## 3. Closed-shell "magic-number" nanocluster

### 3.1 Temperature-induced transition in the spectrum

The majority of cluster ion yield curves display a monotonic post-threshold rise for all temperatures, as illustrated in Fig. 3(a) [43]. These curves can be put to use for extracting the cluster ionization energies (work functions) and their size and temperature variations [53,54], but they do not display any peculiar features. However, we found that for just a few sizes ($Al_{37,44,66,68}$) with decreasing temperature there appears a bulge-like feature close to the ionization threshold. The clearest and most prominent example is observed in the photoionization spectrum of the closed-shell [9,10,46-48] "magic" cluster $Al_{66}$ with 198 valence electrons, as seen in the progression of spectra shown in Fig. 3(b) [43]. We begin by focusing on this cluster. The data for the other sizes will be summarized in the next section.

This emergence of a spectral feature near the top of the electron distribution is a novel effect and the main experimental signature reported here. First of all, it's important to emphasize that it appears only in a few out of the many clusters studied here, and nothing similar is seen in their neighboring sizes. Secondly, while hump-like structures in near-threshold ionization curves have been seen in other nanoclusters with shell structure (e.g., $Cs_n$ and $Cs_nO$ [55,56]) none appeared in closed-shell clusters (such as $Al_{66}$ here) and, most significantly, none were reported to be temperature-dependent. For corroboration, we have measured photoionization yield curves for $Cu_{n=24-87}$ clusters over the same range of temperatures. Those data, reported in [43,53] and with an example shown in Fig. 4, confirm that (i) "magic-number" copper clusters show no notable structure near threshold and (ii) whatever structure is present in some open-shell clusters shows absolutely no significant temperature dependence. Both of these attributes are in strong contrast to what is observed here.

To characterize the evolution of the detected spectra with cluster temperature, we begin by plotting the area under the bulge, see Figs. 3(b) and 5(c). (The plots in Fig. 5 are based on data analysis procedures described in the Appendix and revised as compared with Ref. [43].) This presentation already suggests that an electronic transition is taking place.

### 3.2 Electronic Density of States

Embedded within a photoionization curve there is actually further useful information about the nanoparticle electronic spectrum. Indeed, the photoelectron yield as a function of photon energy $E=\hbar\omega$ is given by

$$Y(E) \propto \int_{-E}^{\infty} M(\varepsilon)\rho_f(E+\varepsilon)f(\varepsilon)D(\varepsilon)d\varepsilon, \qquad (1)$$

where $\varepsilon$ is the electron energy, $M$ the dipole transition matrix element from a shell level into the continuum, $\rho_f$ the DOS of the final (free) electron motion, $f$ the Fermi-Dirac occupation function, and $D$ the electronic DOS within the nanocluster. The energy of the vacuum level is set to zero. Since all the factors but the last one are smooth functions of energy, the derivative $dY/dE$ is essentially proportional to $D(\varepsilon)$. That is, $dY/dE$ provides a direct image of the (temperature-dependent) density of the cluster's electronic states. In a recent paper [53] we confirmed this correspondence by directly superimposing the near-threshold ionization profile derivatives of cold copper clusters onto the corresponding $Cu_n^-$ photoelectron spectra from Ref. [57].

Notice that the situation is to a certain extent analogous to tunneling and scanning-tunneling spectroscopy, where the tunneling current $I$ is given by the convolution of the sample and tip densities of states and the transmission matrix element. Therefore in the first approximation the differential conductance can be written $dI/dV \propto D_S(E_F - eV)$ where $D_S$ is the sample DOS and $E_F$ is the Fermi energy.

By differentiating the $Al_{66}$ nanocluster ionization curves in Fig. 3(b), we find a growing peak in $dY(E)/dE$, shown in Fig. 5(a). Comparing Fig. 5(b) which plots the amplitude of the derivative maximum as a function of cluster temperature, with Fig. 5(c) which plots the area under the bulge, we observe that the plots are similar and point towards an electronic transition taking place at $T \gtrsim 100$ K. More precisely, based on the discussion above we see that the peak shown in Fig. 5(a) is a reflection of the electronic DOS and its change with temperature.

A change in DOS is a well-known signature of the pairing transition. Indeed, in the superconducting scenario the energy spectrum becomes

$$\tilde{\xi} = \left(\xi^2 + \Delta^2\right)^{1/2}, \qquad (2)$$

where $\xi$ is the electron energy in the normal state referred to the chemical potential $\mu$. As a result, the onset of pairing both compresses the highest-occupied electron shell and pushes it downwards [13] (reflecting the extra pair-breaking energy now required to move an electron into the continuum) towards the lower shells which lie quite closely [48]. The consequence is a rise in the near-threshold DOS, as observed.

Such a pattern is familiar from superconductivity in bulk samples, where the DOS has the form [58]

$$D_S = D_0 \frac{\tilde{\xi}}{\sqrt{\tilde{\xi}^2 - \Delta^2}} \Theta(\tilde{\xi} - \Delta). \qquad (3)$$

Here $D_0$ is the DOS at the Fermi level in the normal state, and $\Theta$ the step function.

In Eqs. (2),(3) the order parameter $\Delta$ depends on the temperature. In a finite system the chemical potential also has a temperature dependence because of the requirement of particle number conservation (see, e.g., [13]). The dependence $\Delta(T)$ is especially rapid near $T_c$, which means that the observed change in the photoionization curve likewise takes place near $T_c$.

Once again, it is instructive to draw upon the analogy with the tunneling spectra of superconductors, where gap opening manifests itself via the appearance and growth of prominent lobes in the differential conductance curves, as illustrated in Fig. 6. In our case, since the topmost electrons occupy a shell lying below the vacuum level, the corresponding $dY/dE$ structure is a peak whose intensity grows with decreasing temperature as this shell becomes compressed with the onset of pairing.

### 3.3 Discussion of the Transition

The effect reported here is generally new, and it is natural to interpret it as an electronic transition manifesting superconducting pairing in a nanocluster particle, as described above.

Indeed, the fact that we detect the spectral changes only in a few cluster sizes agrees with the expectation that pairing can take place only in case of propitious combination of electronic degeneracy, shell energies, and coupling strength. Conversely, if this were a structural transition it would be unexpected for it to be so selective in terms of cluster size. Notice also that the temperature of the reported transition lies far below the aluminum clusters' pre-melting and melting points of 300 K - 900 K [60].

Furthermore, the onset of the spectral transition matches the region of theoretically predicted pairing temperatures. According to theory [13], the value of $T_c$ depends sensitively on the occupied-to-unoccupied shell spacing $\delta$ (HOMO-LUMO gap), the degree of degeneracy, and the value of the bulk material's electron-phonon coupling constant $\lambda_b$. $T_c$ is found as the root of a matrix equation which incorporates these factors, accounts for the conservation of particle number, and is not limited to the weak-coupling approximation. It should be emphasized that the effective coupling parameter representing pairing in a finite nanocluster significantly exceeds the corresponding bulk value [13].

The input $\lambda_b$ value should correspond to that for bulk amorphous aluminum, because studies of $Al_n$ clusters suggest that $n=66$ and some other sizes have amorphous-like structure. This characterization is based on density functional calculations of cluster geometries which are supported by experimental measurements of cluster heat capacities [61] and photoelectron spectra [48]. It is known that amorphous materials often display a higher value of $T_c$ than their crystalline counterparts, and correspondingly a higher electron-phonon coupling constant [62,63]. This is indeed the case for amorphous aluminum which, as mentioned above, has $T_c=6$ K [44]. Its value of $\lambda_b$ is unfortunately not tabulated in the literature, but it is known that in many amorphous materials the coupling constant is enhanced by 50% or more, and reaches a value of 2 or even higher [62,63]. For example, bulk amorphous Ga, an element chemically similar to aluminum, has $\lambda_b=1.9$-$2.25$ [62].

Taking, therefore, for an estimate $\lambda_b \approx 2$ together with the $Al_{66}$ intershell spacing of $\delta \approx 0.3$-$0.35$ eV (deduced from photoelectron spectroscopy data [10,48] as the distance between the half-maximum points on the facing slopes of the topmost peaks), so that $\xi$ in Eq. (2) is $\approx 0.15$-$0.17$ eV (since $\mu$ is located halfway between the highest-occupied and lowest-unoccupied shells [13]), and a characteristic vibrational frequency $\tilde{\Omega} \sim 35$ meV, it is found that the equation for $T_c$ yields a solution of $\approx 70$ K [64]. Considering the degree of uncertainty in the numerical parameters (indeed, $\lambda_b$, $\tilde{\Omega}$, the Fermi momentum, and the matrix elements are all deduced from bulk measurements and may have somewhat different values in nanoclusters), this is in very sensible agreement with the location of the spectral transition observed here. The corresponding magnitude of the energy gap is estimated as $2\Delta \sim 0.1$ eV [64] (recall that in the strong-coupling regime $2\Delta/T_c$ can markedly exceed the 3.52 weak-coupling BCS ratio [14]). This is commensurate with aforementioned value of $\xi$, and is thus consistent both with the criterion for pairing correlations being observable [65] and with the position and width of the experimentally observed bulge (Figs. 3,5(a)).

Finally, note that the gradual decrease in the intensity of the bulge above the transition may reflect pairing fluctuations expected in a finite system [66,67]. At the same time, as mentioned above, order parameter fluctuations will not extinguish the transition: under the present strong pairing conditions the superconducting coherence length remains comparable to the cluster size.

### 4. "Non-magic" clusters

For nanocluster sizes which do not possess a filled spherically symmetric electronic shell, the state degeneracy is lower, but the chemical potential $\mu$ becomes positioned at the highest occupied electronic levels (as opposed to lying halfway up to the next unoccupied shell as in the closed-shell "magic" cluster discussed above) and the level spacing also decreases because of the geometrical distortion of the cluster [48]. The first effect works against pairing but the last two make it more favorable, hence high $T_c$ may occur in open-shell clusters as well.

As mentioned above, in addition to the clear observation of a transition in the "magic" $Al_{66}$ cluster, inspection of the photoionization curves also revealed the appearance of "bulges" with decreasing temperature in the spectra of $Al_{37}$, $Al_{44}$, and $Al_{68}$. The data, and their treatment along the same lines as Figs. 3,5 are presented in Figs. 7-9. The data for these cluster sizes had more scatter than for $Al_{66}$, hence the error bars are much higher and the transition and $dY/dE$ curves are not as robust (we will be undertaking further measurements to map out the spectra with higher precision). Nevertheless, there is strong qualitative evidence for a similar temperature-induced transition in the density of states [68].

It is interesting to note that all three of these cluster sizes are located near (or at) electronic shell closings at $Al_{36}, Al_{44}, Al_{66}$ (for the first two the closings are assisted by their geometric packing [48]), a situation identified theoretically as favorable for pairing [13].

## 5. Conclusions

This paper summarizes our measurements of the photoionization spectra of free, size-selected, "superatom" aluminum nanoclusters, in which the presence of discrete electronic shell structure turns out to be very favorable for the possibility of extremely strong pairing. By means of such spectroscopy we were able to obtain a view of the temperature-dependent density of states of the topmost (near the Fermi level) cluster electrons. In four clusters in the studied size range ($Al_{37,44,66,68}$) the data revealed a novel feature – a "bulge" appearing near the threshold of the spectrum and rising dramatically as the cluster temperature was lowered towards ~100 K. As discussed above, this phenomenon, previously unobserved, is consistent in every way with the predicted pairing transition. This holds the promise of the appearance of a completely new class of high-temperature superconductors, which may be extended to still much higher critical temperatures by the optimization of size, material, and composition.

In future work, we will enhance the temperature resolution and range, and explore further sizes and materials, including Zn, Cd, Ga and In which have all been raised as possible "superconducting superatom" candidates, as well as mixed (i.e., alloyed) nanoclusters which are easily attainable with today's cluster-beam sources.

It is noteworthy that although photoelectron current measurements are not frequently applied to bulk superconductors, Refs. [69,70] did observe that near-threshold photoelectron yield from the surface of $Bi_2Sr_2CaCu_2O_{8+\delta}$ undergoes marked changes at the transition point. In nice resemblance to cluster behavior described above, the cuprate photoyield spectrum acquired new structures that could be ascribed to changes in the electronic state density. It was also found that the total amount of cuprate photocurrent was noticeably different above and below $T_c$. Such an absolute yield measurement is presently inaccessible to free-cluster experiments because the neutral cluster flux in the beam is itself affected by the thermalizing tube temperature, but it would be very interesting to pursue.

In complex materials with nontrivial phase diagrams, one can observe the appearance of a "pseudogap" in the electronic spectrum which is distinct from the pairing gap (see, e.g., [71]). However, there is no such behavior in the simple aluminum metal and so it would also be unlikely for the observed spectral transition to reflect some kind of pseudogap phenomenon. The main distinctive aspect of a nanocluster lies in the discreteness of its electronic spectrum and not in the appearance of new complex phases. Indeed, as emphasized above, the observed transition is fully consistent with the theoretically predicted onset of high-temperature pairing. Of course a direct study of the resulting coherence of the electronic state would also be valuable, although challenging to implement in a beam experiment. One possible technique would be a search for angular and momentum correlations in two-electron emission spectroscopy [72-74].

A related question concerns the influence of a magnetic field. Unfortunately a direct pursuit of the Meissner effect is difficult, as remarked in Sec. I. Furthermore, setting $\left(H_c^2/8\pi\right)V \approx N\Delta$ (here $V$ is the particle volume and $N$ is the number of paired electrons in the uppermost shell), one finds for the critical field $H_c$~10 T, which is presently impractical for mass spectrometers or cluster beam machines in general. Possible options may involve using cold cluster ion traps (see, e.g., [75-77]) adapted to very high magnetic fields, or - when the transition is fully mapped out - working with weaker fields very close to $T_c$.

While accurate size-selective measurements on individual free clusters are essential for identifying and characterizing this novel superconducting family, future applications will require, and make use of, assembling such superconducting size-selected nanoclusters into arrays, films, and compounds. Consider, for example, a chain or network made up of identical nanoclusters with discrete shell-ordered energy spectra, connected by tunneling barriers. Recent theory predicts that such a chain is capable of supporting Josephson tunneling current two to three orders of magnitude stronger than in conventional systems [78]. Thus the use of high-$T_c$ nanoclusters could combine an orders-of-magnitude increase in superconducting current capacity with an orders-of-magnitude increase in the operating temperature.

A number of actively researched approaches have the potential to reach this goal. While not yet demonstrated, soft-landing of an array of identical nanoclusters with shell structure on a surface template should become realistic at some point. In addition, the synthesis of ordered crystals [79,80] out of identical ligand-protected clusters represents a very promising course. In some compounds of this type the metal core retains shell structure ordering, while the outer protective shell may be able to provide the tunneling barrier. In fact, a $Ga_{84}$-cluster compound has been shown to exhibit superconductivity with $T_c \approx 8$ K, a seven-fold increase over the critical temperature of bulk gallium [81,82]. The work on discovery and characterization of high-$T_c$ pairing in individual nanoclusters, as introduced in the present paper, is therefore valuable both for the inherently novel physics and for the identification of promising building blocks for such new materials.

**Acknowledgement**
We would like to thank Dr. V. Z. Kresin and Dr. Yu. N. Ovchinnikov for useful discussions, Dr. Anthony Liang for his collaboration at the beginning of this work, and Dr. Chunrong Yin for the construction of the nanocluster source. We are grateful to the staff of the USC Machine Shop for their invaluable contributions to the project, and to the engineers at Altos Photonics for their help in maintaining the performance of the tunable laser. This research is supported by the U. S. National Science Foundation under Grant No. DMR–1206334.

**Appendix**
*Evaluation of the area plots.* Ionization spectra from different runs for a specific cluster size and temperature were interpolated in 10 meV segments by smoothing and cubic spline fitting. With the data thus cast in the form of an array $Y_m$ corresponding to photon energies $E_m$ evenly spaced by 10 meV, the area under the bulge is proportional to $\sum_m A_m \equiv \sum_i \left(Y_m - Y_m^{lin}\right)$ where $Y_i^{lin}$ is the underlying straight dashed line in Fig. 3(b) drawn between endpoints $(E_i, Y_i)$ and $(E_f, Y_f)$. Therefore the error bar for the total area is composed, in quadrature, of those for individual points, $A_m = Y_m - \left[Y_i + (Y_f - Y_i)(E_m - E_i)/(E_f - E_i)\right]$ which are calculated via propagation-of-error formulas. The standard deviations of $Y_m$ are calculated from scatter between individual runs, and those of $Y_i$ and $Y_f$ are approximated by the average over all $m$ points.

*Evaluation of the amplitude ratio plots.* The spline-fitted data sets were differentiated and smoothed. In the plots, the array $d_m \equiv (dY/dE)_m$ is spaced by 10 meV intervals and normalized to the height $d_{min}$ of the derivative minimum that follows the peak $d_{max}$. That is, Figs. 5(a,b) show $d_m/d_{min}$ and $d_{max}/d_{min}$, respectively. The standard deviations, $\sigma_{d_m}$, of $d_m$ values is found from the variation between curves from individual runs. To find the optimal fits for the peak and the valley while ensuring that they are not excessively skewed by some individual data points, we created 100 synthetic profiles out of points $d_m + R\sigma_{d_m}$, where $R$ is a normally distributed random number, within an energy range of ±30-50 meV around $E_{max}$ and $E_{min}$. $d_{max}$ and $d_{min}$ values and their standard deviations were derived from Gaussian or quadratic fits to these sets of profiles, and used to calculate the points and error bars in Figs. 5(a,b).

**Figures**

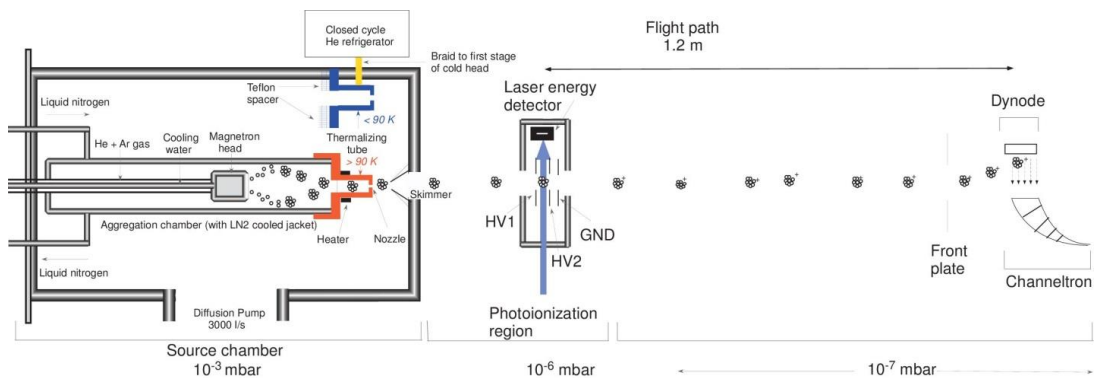

**Fig. 1**. Scheme of the experimental setup (not to scale). The source produces a flux of neutral aluminum nanocluster particles by ion sputtering followed by aggregation growth. The clusters are thermalized to the desired internal temperature by passing through a thermalizing tube mounted to the end face of the aggregation zone. Depending on the desired temperature, this tube either holds a band heater or is connected to a refrigerator cold head. The clusters are then ionized by a pulsed tunable laser and extracted into a time-of-flight mass spectrometer. In this way their ionization spectra can be mapped as a function of size, temperature, and wavelength.

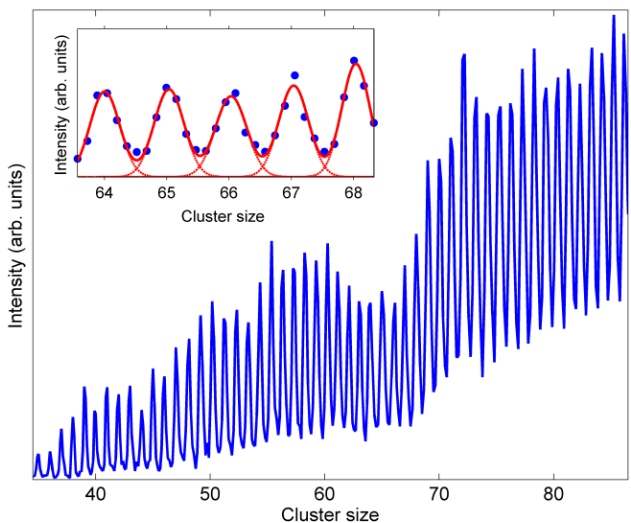

**Fig.2.** An example of an $Al_n$ time-of-flight mass spectrum. Inset: deconvolution of mass spectral intensities. See Sec. 2 for details.

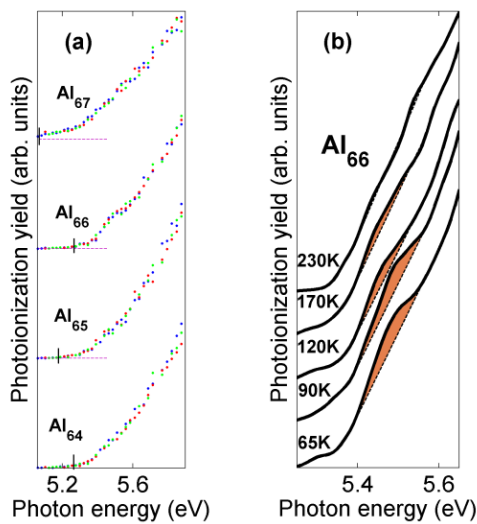

**Fig. 3.** (a) Photoionization yield plots for several $Al_n$ nanoclusters obtained at $T$=65K. The curves are shown shifted with respect to each other for clarity. Short vertical bars denote the cluster ionization threshold energies. A strong bulge-like feature appears close to the threshold for $n$=66. The adjacent clusters show no such feature. The sharp drop in the ionization energy from $Al_{66}$ to $Al_{67}$ reflects the fact that the former is a "superatom" with a filled electronic shell. Different color dots correspond to data from several experimental runs. (b) The strengthening of the $Al_{66}$ spectral feature with decreasing temperature can be seen by comparing the thick experimental yield curve (a spline average of the data from repeated runs) with the dashed interpolating line. (This figure also appeared as Fig. 1 of Ref. [43].)

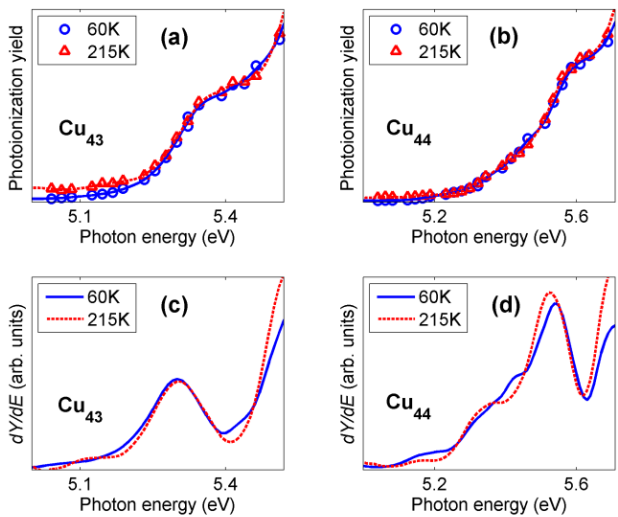

**Fig. 4**. Photoionization yield curves of copper nanoclusters, illustrated here for a pair of representative sizes (a,b) together with their derivatives (c,d), show no temperature-dependent features. This supports the conclusion that the aluminum data in Figs. 3,7-9 reveal a distinct electronic transition.

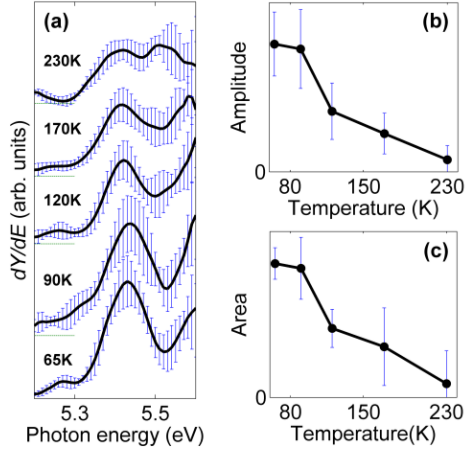

**Fig. 5.** Temperature dependence of the $Al_{66}$ spectrum and the density of states. (a) Derivatives of the near-threshold portion of the photoionization yield plots from Fig. 3(b). As discussed in the text, $dY/dE$ represents a measure of the electronic density of states. The intensity of the first peak, which derives from the "bulge" in the $Al_{66}$ spectrum, grows with decreasing temperature, implying a rise in the density of states near threshold. The plots are normalized to the amplitude height of the minimum following the derivative peak. (b) To quantify the intensity variation of the peak in (a), we plot its amplitude as a function of cluster temperature. (c) Another measure of the magnitude of the bulge is its area relative to the dashed straight line in Fig. 3(b). It is noteworthy that the behavior of the plots in panels (b) and (c) matches, both suggesting that a transition takes place as the temperature approaches ≈100 K.

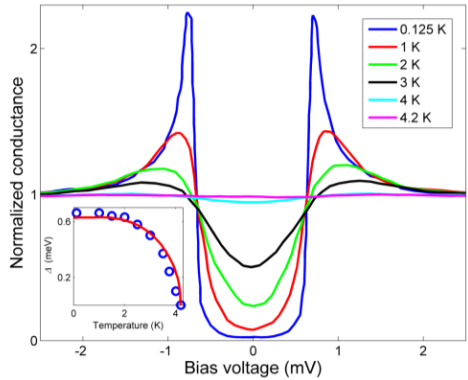

**Fig. 6.** A plot of scanning tunneling spectroscopy data for superconducting amorphous tungsten-based nanoscale deposits ($T_c$=4.15 K), after Ref. [59]. The purpose of showing this plot is to highlight the physical similarity between photoemission yield spectroscopy employed in the present work and conductance spectra from tunneling experiments. As discussed in Sec. 3.2, in both cases the curves reflect the electronic density of states of the sample near the Fermi level. The appearance and growth of the lobe in the differential conductance (due to the opening of the superconducting gap in the continuous electronic spectrum, Eq. (3)) is analogous to the growth of the bulge in the differential photoyield ($dY/dE$) shown in Fig. 5(a) (due to the compression of the density of states within a discrete electronic shell). One difference to note is that tunneling can proceed in both directions, hence the conductance curve displays two lobes, while cluster photoionization is unidirectional and gives rise to only a single peak. The dependence of the pairing gap on temperature (inset) and therefore the variation of the spectrum are especially rapid near $T_c$.

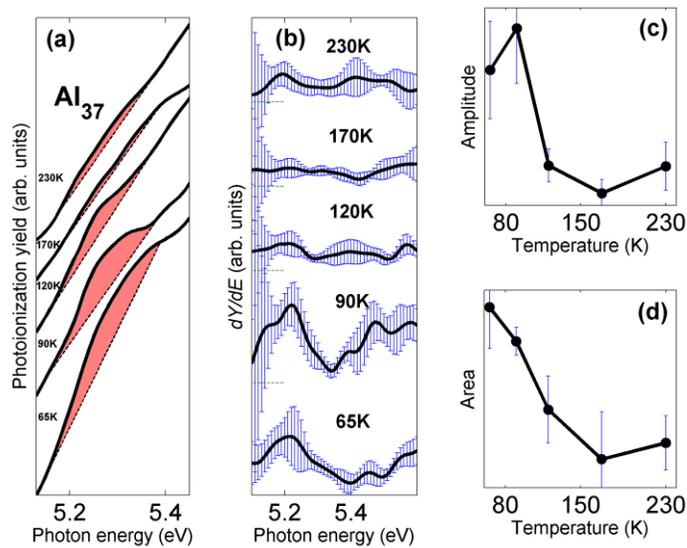

**Fig. 7.** Al$_{37}$ ionization spectra. (a) Photoionization yield "bulge" and (b) its corresponding normalized derivatives, presented similarly to Figs. 3(b) and 5(a). (c,d) The temperature dependence of the derivative peak amplitude and of the bulge area relative to the interpolating line, as in Figs. 5(b,c).

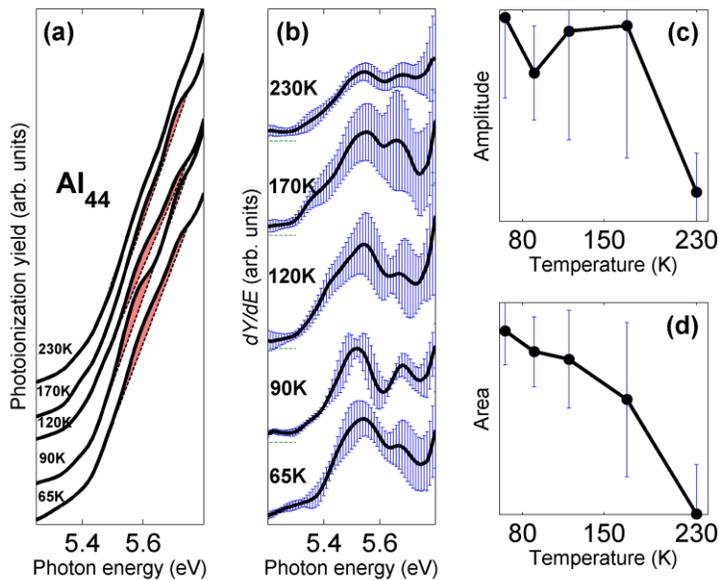

**Fig. 8.** Al$_{44}$ ionization spectra, presented as in Fig. 7.

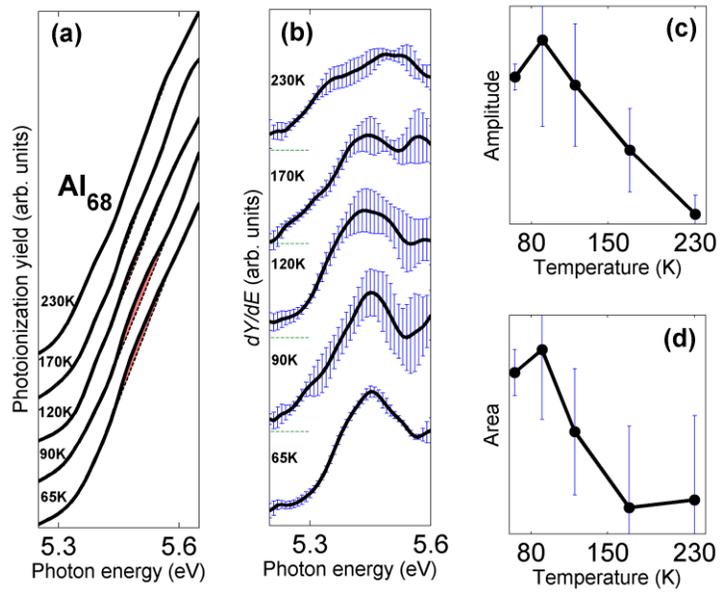

**Fig. 9.** Al$_{68}$ ionization spectra, presented as in Fig. 7.

# References

1. Johnston, R. L.: Atomic and Molecular Clusters. Taylor&Francis, London (2002)
2. Haberland. H. (ed.): Clusters of Atoms and Molecules. Springer, Berlin (1994)
3. Alonso, J. A.: Structure and Properties of Atomic Nanoclusters. 2nd ed. Imperial College Press, London (2012)
4. de Heer, W.: Rev. Mod. Phys. **65**, 611 (1993)
5. Knight, W. D., Clemenger, K., de Heer, W. A., Saunders, W. A., Chou, M. Y., Cohen, M. L.: Phys. Rev. Lett. **52**, 2141 (1984)
6. Castleman, Jr., A. W., Bowen, Jr., K. H.: J. Phys. Chem. **100**, 12911 (1996)
7. v. Issendorff, B. In: Sattler, K. D. (ed) Handbook of Nanophysics: Clusters and Fullerenes, Vol. 2, Chap. 6. CRC Press, Boca Raton (2011)
8. Khanna, S., Jena, P.: Phys. Rev. B **51**, 13705 (1995)
9. Persson, J. L., Whetten, R. L., Cheng, H.-P., Berry, R. S.: Chem. Phys. Lett. **186**, 215 (1991)
10. Li, X., Wu, H., Wang, X.-B., Wang, L.-S.: Phys. Rev. Lett. **81,** 1909 (1998)
11. Analogous (but with lower degeneracy) DOS peaks arise in the study of enhanced superconductivity in quantized thin films and nanowires, see, e.g., the review [12].
12. Peeters, F. M., Shanenko, A. A., Croitoru, M. D. In: Sattler, K. D. (ed) Handbook of Nanophysics: Principles and Methods, Vol. 1, Chap. 9. CRC Press, Boca Raton (2011)
13. Kresin, V. Z., Ovchinnikov, Yu. N.: Phys. Rev. B **74**, 024514 (2006)
14. Kresin, V. Z., Morawitz, H., Wolf, S. A.: Superconducting State: Mechanisms and Properties. Oxford University, Oxford (2014)
15. Bohr, A., Mottelson, B. R., Pines, D.: Phys. Rev. **110**, 936 (1958)
16. Ring, P., Schuck, P.: The Nuclear Many-Body Problem. Springer, Berlin (1980)
17. Broglia, R. A., Zelevinsky, V. (eds.): Fifty Years of Nuclear BCS: Pairing in Finite Systems. World, Singapore (2013)
18. Heiselberg, H., Mottelson, B.: Phys. Rev. Lett. **88**, 190401 (2002)
19. Regal, C. A., Jin, D. S. In: Berman, P. R., Lin, C. C., Arimondo, E. (eds.) Advances in Atomic, Molecular, and Optical Physics, Vol. 54, pp. 1-79. Academic, Waltham (2007)
20. Kresin, V. Z., Litovchenko, V. A., and Panasenko, A. G.: J. Chem. Phys. **63**, 3613 (1975)
21. Clark, K., Hassanien, A., Khan, S., Braun, K.-F., Tanaka, H., Hla, S.-W.: Nature Nanotech. **5**, 261 (2010)
22. Cohen, R. W., Abeles, B.: Phys. Rev. **168**, 444 (1968)
23. Hauser, J. J.: Phys. Rev. B **3**, 1611(1971)
24. Deutscher, G., Gershenson, M., Grünbaum, E., Imry, Y.: J. Vac. Sci. Tech. **10**, 697 (1973)
25. Leemann, C., Elliott, J. H., Deutscher, G., Orbach, R., Wolf, S. A.: Phys. Rev. B **28**, 1644 (1983)

26. Abeles, P. In: Wolfe, R. (ed.) Applied Solid State Science, Vol. 6, pp. 1-117. Academic, New York (1976)

27. Perenboom, J. A. A. J., Wyder, P., Meier, F.: Phys. Rep. **78**, 174 (1981)

28. Von Delft, J., Ralph, D. C.: Phys. Rep. **345**, 61 (2001)

29. Bose, S., Ayyub, P.: Rep. Prog. Phys. **77**, 116503 (2014)

30. Friedel, J.: J. Phys. II France **2**, 959 (1992)

31. Knight, W. D. In: Wolf, S. A., Kresin, V. Z. (eds) Novel Superconductivity. Plenum, New York (1987)

32. Mottelson, B.: Nucl. Phys. A **574**, 365 (1994)

33. Lindenfeld, Z., Eisenberg, E., Lifshitz, R.: Phys. Rev. B **84**, 064532 (2011)

34. Croitoru, M. D., Shanenko, A. A., Kaun, C. C., Peeters, F. M.: Phys. Rev. B **83**, 214509 (2011)

35. Baturin, V. S., Losyakov, V. V.: J. Exp. Theor. Phys. **112**, 226 (2011)

36. de Heer, W. A., Kresin, V.V. In: Sattler, K. D. (ed) Handbook of Nanophysics: Clusters and Fullerenes, Vol. 2, Chap. 10. CRC Press, Boca Raton (2011)

37. Peculiar odd-even effects in the electric susceptibilities of cold (~20 K) niobium nanoclusters have been hypothesized [38] to result from pairing. These interesting effects deserve more exploration, but Nb clusters do not exhibit electronic shell structure and therefore lie outside the cluster families discussed here. Ref. [39] observed small heat capacity jumps at T≈200 K for a pair of free aluminum cluster ions and suggested that they may be consistent with pairing, however the data statistics were limited.

38. Yin, S., Xu, X., Liang, A., Bowlan, J., Moro, R., de Heer, W. A.: J. Supercond. Nov. Magn. **21**, 265 (2008)

39. Cao, B., Neal, C. M., Starace, A. K., Ovchinnikov, Y. N., Kresin, V. Z., Jarrold, M. F.: J. Supercond. Nov. Magn. **21**, 163 (2008)

40. Kresin, V. V.: Phys. Rev. B **38**, 3741 (1988)

41. van Ruitenbeek, J. M., van Leeuwen, D. A.: Phys. Rev. Lett. **67**, 640 (1991)

42. Roduner, E., Jensen, C., van Slageren, J., Rakoczy, R. A., Larlus, O., Hunger, M.: Angew. Chem. Int. Edit. **53**, 4318 (2014)

43. Halder, A., Liang, A., Kresin, V. V.: Nano Lett. **15**, 1410 (2015)

44. Tsuei, C. C. In: Buschow, K. H. J. (ed) Concise Encyclopedia of Magnetic and Superconducting Materials, 2nd ed.. Elsevier, Amsterdam (2005)

45. de Heer, W., Milani, P., Châtelain, A.: Phys. Rev. Lett. **63**, 2834 (1989)

46. Pellarin, M., Baguenard, B., Broyer, M., Lermé, J., Vialle, J. L., Perez, A: J. Chem. Phys. **98**, 944 (1993)

47. Schriver, K. E., Persson, J. L., Honea, E. C., Whetten, R. L.: Phys. Rev. Lett. **64**, 2539 (1990).

48. Ma, L., von Issendorff, B., Aguado, A.: J. Chem. Phys. **132**, 104303 (2010)

49. Haberland, H., Karrais, M., Mall, M., Thurner, Y.: J. Vac. Sci. Technol. A **10**, 3266 (1992)

50. Haberland, H., Mall, M., Moseler, M., Qiang, Y., Reiners, T., Thurner, Y.: J. Vac. Sci. Technol. A **12**, 2925 (1994)

51. Halder, A.: Ph. D. thesis. University of Southern California, Los Angeles (2015)

52. Hock, C., Schmidt, M., von Issendorff, B.: Phys. Rev. B **84**, 113401 (2011)

53. Halder, A., Huang, C., Kresin, V. V.: J. Phys. Chem. C **119**, 11178 (2015)

54. Halder, A., Kresin, V. V.: J. Chem. Phys. **143**, 164313 (2015)

55. Bergmann, T., Martin, T. P.: J. Chem. Phys. **90**, 2848 (1989)

56. Limberger, H. G., Martin, T. P.: J. Chem. Phys. **90**, 2979 (1989)

57. Kostko, O.: Ph. D. thesis. Albert-Ludwigs-Universität, Freiburg (2007)

58. Tinkham, M.: Introduction to Superconductivity. 2nd ed. McGraw-Hill, New York, 1996

59. Guillamón, I., Suderow, H., Vieira, S., Fernández-Pacheco, A., Sesé, J., Córdoba, R., De Teresa, J. M., Ibarra, M. R., New J. Phys. **10**, 093005 (2008)

60. Aguado, A., Jarrold, M. F.: Annu. Rev. Phys. Chem. **62**, 151 (2011)

61. Starace, A. K., Neal, C. M., Cao, B., Jarrold, M. F., Aguado, A., Lopez, J. M.: J. Chem. Phys. **131**, 044307 (2009)

62. Bergmann, G.: Phys. Rep. **27**, 159 (1976)

63. Grimvall. G.: The Electron-Phonon Interaction in Metals. North-Holland, Amsterdam, 1981.

64. Yu. N. Ovchinnikov and V. Z. Kresin, private communication

65. Anderson, P. W.: J. Phys. Chem. Solids **11**, 26 (1959)

66. Mühlschlegel, B., Scalapino, D. J., Denton, B.: Phys. Rev. B. **6**, 1767 (1972)

67. Larkin, A., Varlamov, A.: Theory of Fluctuations in Superconductors. Oxford University Press, Oxford (2005)

68. $Al_{37}$ has an odd number of electrons, but the same scenario can nevertheless take place here: the near-threshold photoemission curve derives from one unpaired electron, plus a bulge due to the much larger number of the other paired electrons.

69. Park, W. G., Nepijko, S. A., Fanelsa, A., Kisker, E., Winkeler, L., Solid State Commun. **91**, 655 (1994)

70. Nepijko, S. A., Park, W. G., Fanelsa, A., Kisker, E., Winkeler, L., Güntherodt, G.: Physica C **288**, 173 (1997)

71. Müller, K. A., Bussmann-Holder, A. (eds.): Superconductivity in Complex Systems. Springer, Berlin (2010)

72. Berakdar, J., Kirschner, J. (eds.): Correlation Spectroscopy of Surfaces, Thin Films, and Nanostructures. Wiley, Weinheim (2004).

73. Becker, U., Shirley, D. A. (eds.): VUV and Soft X-Ray Photoionization. Plenum, New York (1996).

74. Wehlitz, R., Juranić, P. N., Collins, K., Reilly, B., Makoutz, E., Hartman, T., Appathurai, N., Whitfield, S. B.: Phys. Rev. Lett. **109**, 193001 (2012).

75. Terasaki, A, Majima, T., Kondow, T.: J. Chem. Phys. **127**, 231101 (2007)

76. Peredkov, S., Neeb, M., Eberhardt, W., Meyer, J., Tombers, M., Kampschulte, H., Niedner-Schatteburg, G.: Phys. Rev. Lett. **107**, 233401 (2011)

77. Martinez, F., Bandelow, S., Breitenfeldt, C., Marx, G., Schweikhard, L., Vass, A., Wienholtz, F.: Int.. J. Mass Spectrom. **365-366**, 266 (2014)

78. Ovchinnikov, Y. N., Kresin, V. Z.: Phys. Rev. B **81**, 214505 (2010)

79. Weitz, I. S., Sample, J. L., Ries, R., Spain, E. M., Heath, J. R.: J. Phys. Chem. B **104**, 4288 (2000)

80. Häkkinen, H., Tsukuda, T. (eds.): Protected Metal Clusters: From Fundamentals To Applications. Elsevier, Amsterdam (2015)

81. Bakharev, O. N., Bono, D., Brom, H. B., Schnepf, A., Schnöckel, H., de Jongh L. J.: Phys. Rev. Lett. **96**, 117002 (2006)

82. Bono, D., Schnepf, A., Hartig, J., Schnöckel, H., Nieuwenhuys, G. J., Amato, A., de Jongh, L. J.: Phys. Rev. Lett. **97**, 077601 (2006)